\documentclass[a4paper,11pt]{article}

\usepackage[english]{babel}
\usepackage[utf8]{inputenc}
\usepackage{lmodern} 
\usepackage{here}
\usepackage{float}
\usepackage{textcomp}
\usepackage{enumitem}
\usepackage{here}
\usepackage[square,sort]{natbib}
\usepackage{wrapfig}
\usepackage{colortbl}
\usepackage{tabu}
\usepackage{algpseudocode}
\usepackage{varwidth}
\usepackage{tabularx}
\usepackage[section]{placeins}
\usepackage{slashbox}
\usepackage{multirow}
\usepackage{indentfirst}
\usepackage[nottoc,notlot,notlof]{tocbibind}
\usepackage{algorithm,algorithmicx}
\usepackage{authblk}

\usepackage{graphicx} 

\RequirePackage{etoolbox}
\csdef{input@path}{%
 {sty/}
 {img/}
}%
\csgdef{bibdir}{bib/}

\usepackage{amsmath}
\usepackage{amsthm}
\usepackage{amsfonts}
\usepackage{hyperref}

\usepackage[margin=1in,headheight=13.6pt]{geometry}

\usepackage{pgf,tikz}
\usepackage{pifont}

	\usetikzlibrary{arrows,shapes,positioning}
\newsavebox{\tempbox}

\newcommand{\textbox}[1]
{\savebox{\tempbox}{#1}
 \ifdim\wd\tempbox<4cm\relax
   \makebox[4cm]{\usebox{\tempbox}}%
 \else
   \parbox{4cm}{\raggedright #1}%
 \fi}

\theoremstyle{plain}
\usetikzlibrary{positioning}

\tikzset{basic/.style = {draw,text width=2cm, rectangle,align=center}}
\tikzset{trait/.style = {line width=2pt}}
\tikzset{fleche/.style = {trait,>=stealth,->}}

 \usepackage[automake,acronym,nonumberlist,toc,section=chapter]{glossaries}

\makeglossaries

\newacronym{gr}{GR}{Gamma Ray, unit: API}
\newacronym{phin}{$\phi_N$}{Neutron porosity, unit: $V/V$}
\newacronym{rhob}{$\rho_b$}{Bulk Density, unit: $g.cm^{-3}$}
\newacronym{pef}{$pef$}{Photoelectric factor, unit: $barns/electron$}
\newacronym{sonic}{$\Delta_t$}{Sonic, unit: $US/ft$}
\newacronym{res}{$R_t$}{Resistivity, unit: $\Omega.m$}
\newacronym{sp}{SP}{Spontaneous potential, unit: mV}
\newacronym{nmr}{NMR}{Nuclear magnetic resonance}
\newacronym{DRHO}{DRHO}{Bulk density correction, unit: $g.cm^{-3}$}
\newacronym{MD}{MD}{Measured depth in ft}
\newacronym{TVD}{TVD}{True vertical depth in ft}
\newacronym{PELT}{PELT}{Pruned exact linear time}
\newacronym{BinSeg}{BinSeg}{Binary Segmentation}
\newacronym{DTW}{DTW}{Dynamic time warping}
\newacronym{MSA}{MSA}{Multiple sequence alignment}
\newacronym{SAX}{SAX}{Symbolic Aggregate approXimation}
\newacronym{PAA}{PAA}{Piecewise aggregate approximation}
\newacronym{BLOSUM}{BLOSUM}{ BLOcks SUbstitution Matrix}
\newacronym{PAM}{PAM}{ Point accepted mutation}
\newacronym{t-SNE}{t-SNE}{t-distributed stochastic neighbor embedding}
\newacronym{ABC}{ABC}{Approximate Bayesian Computation}
\newacronym{HDBSCAN}{HDBSCAN}{ Hierarchical density-based spatial clustering of applications with noise}
\newacronym{PSO}{PSO}{Particle swarm optimization}

\begin{document}

\title{Evaluation of mineralogy per geological layers by Approximate Bayesian Computation}

\author[1]{Vianney Bruned}
\author[2]{Alice Cleynen}
\author[2]{André Mas}
\author[1]{Sylvain Wlodarczyck}

\affil[1]{Schlumberger Petroleum Services, Montpellier, 34000, France.}
\affil[2]{IMAG, Univ Montpellier, CNRS, Montpellier, France.}



\date{}

\maketitle

\begin{abstract}
We propose a new methodology to perform mineralogic inversion from wellbore logs based on a Bayesian linear regression model. Our method essentially relies on three steps. The first step makes use of Approximate Bayesian Computation (ABC) and selects from the Bayesian generator a set of candidates-volumes corresponding closely to the wellbore data responses. The second step gathers these candidates through a density-based clustering algorithm. A mineral scenario is assigned to each cluster through direct mineralogical inversion, and we provide a confidence estimate for each lithological hypothesis. The advantage of this approach is to explore all possible mineralogy hypotheses that match the wellbore data. This pipeline is tested on both synthetic and real datasets.
\end{abstract}
Keywords : Mineralogical Inversion; Inverse problem; Wellbore log; Approximate Bayesian Computation; Clustering

\section{Introduction}
One of the main goals of reservoir evaluation is the determination of petrophysical parameters like porosity, permeability or water saturation. In order to get an accurate estimation of these parameters, a complete characterization of the lithology or the nature of the rocks is necessary. The petrophysicist proceeds to the analysis of wellbore logs which often requires the input from an expert. Indeed, petrophysical inversion of wellbore logs yields a selection of minerals or fluids belonging to the formation usually with more unknowns (the mineralogy) than measurements (the logs). In a bulk density-neutron porosity cross-plot, an expert may identify the presence of gas, limestone or an exotic mineral. But these choices may not always be obvious from
a direct lecture of the logs.

We can summarize roughly the characterization of the lithology based on conventional logs using a classical inversion approach like Elan or Multi-min  (described in \cite{mayer1980global}, \cite{quirein1986coherent}, \cite{cannon1990applying} or \cite{peeters1991comparison}) by the following steps:
\begin{itemize}
\item Definition of the zones of the well.
\item Selection of the mineralogical components and fluids for each zone, and computation of the lithology using an inversion approach.
\item Tuning the physical parameters (endpoints) of the components (when needed).
\end{itemize}

The petrophysicist will iterate the last two points until they reach a model fitting the data or having a good match with core data. Tuning the endpoints is often necessary for the shaly component. For instance, the variability of the endpoints of the kerogen is very high through the geographical areas: the gamma ray ranges between 500 API to 4000 API. 

We propose a Bayesian approach to select minerals in a stratum. Forward modeling liberates us from the underdetermination of the classical inversion problem using conventional logs. Bayesian inversion methods have been mainly used to solve inverse problems in the domain of rock-physics or geophysics (\cite{tarantola2005inverse}). Many methods use a Bayesian framework to solve the amplitude versus offset (AVO) inversion and to obtain petrophysical attributes like porosity, water saturation or volume of clay (\cite{grana2016bayesian}, \cite{xu2016bayesian}, \cite{rimstad2012hierarchical}).

Other petrophysical approaches to retrieve interesting features have been tested in \cite{qin2017bayesian}, \cite{da2008automatic}, \cite{sanchez2010synthetic}. These methods mainly focus on the uncertainties of the logs and the analysis of the posterior distribution of the petrophysical outputs. Except in \cite{yang2013joint}, the minerals considered in the Bayesian inversion are very limited  (Shale, Sand, Water, Hydrocarbron and sometimes Carbonate) and the model linking the minerals to the logs are quite simple. Moreover, these methods often have a number of unknowns equal to the number of logs available.

In this paper, we assume that the geological layers are known (we may refer to historical references \cite{Wolf1982}, \cite{moline1991identification} and \cite{NewFacies} or to the more recent \cite{Facies3} for a description of some segmentation algorithms for layer detection). Hence we assume that these geological layers have a constant mineral composition, and we describe the model in Section \ref{model.context}. Unlike previous works, our model is flexible enough to allow complex relationships between the logs and the volumes, and we do not restrict ourselves to a low number of minerals. We introduce a three-step method to get the different lithological hypotheses per layers in order to solve the underdetermined mineralogical inversion problem. First, we use an Approximate Bayesian Computation (ABC) technique to get the posterior probability of the lithology (see Section \ref{method}). Then we use a density-based clustering algorithm on this posterior probability to distinguish the different lithological hypotheses, as described in Section \ref{clustering}. Finally, for each of these hypotheses, we propose a method to tune the endpoints based on the resolution of a global optimization problem. 
Our method is therefore entirely automated, and proposes different plausible lithological hypotheses as well as some confidence estimate for each of them. We illustrate its performance on both synthetic (see Section \ref{synth.data}) and real datasets (see Section \ref{case.study}.)

\section{Methodology }
\subsection{Model and Context} \label{model.context}

In mineralogical inversion, a crucial step is the choice of the components of the lithology. Indeed,  starting from $d$ logs and using a classical inversion imply a maximum of $d+1$ minerals or fluids in the inversion model. But usually, $M$ the number of mineral components is such that $M \gg d$, so a model selection is needed. Here, we denote the elements of the lithology by $V \in  \mathbb{R}^{M}$ or volume (volumetric fraction) and the logs by $L \in  \mathbb{R}^{d} $.  $V$ represents the volumetric fraction of the minerals and the fluids. We assume that for each depth $n$, we have: 
\begin{equation}
L_{n}=G\left( V_{n}\right)+X_{n},
\label{eq:log_equation}
\end{equation}
where $X_{n}\sim N\left(  0,\Sigma\right) $ ($\mathcal{N}$ is a multivariate normal distribution and $\Sigma \in \mathcal{M}\left( \mathbb{R}^{d} \right) $ is the covariance matrix)  and $G$ an operator from $\mathbb{R}^{M}$ to $\mathbb{R}^{d}$. Besides, the volumes $V_{i,n}$ are constrained by:
\begin{align*}
\sum_{i=1}^{M} V_{i,n} &=1 , \\
V_{i,n}  \geq 0 & \text{ for all i}.   
\end{align*}
Our aim is to select the minerals that may appear within the geological stratum. Computing the exact amount of the volumetric fraction of the minerals is performed in the final step. Notice that whenever the number of selected components is $M$ with $M<d+1$, a classical inversion program can be run to obtain the exact volumetric fraction. The physical parameters/endpoints of the different minerals or fluids are fixed.

We consider that $d$ is around 4-5, we often have a triple combo (gamma ray: GR, neutron porosity: $\phi_{N}$, bulk density: $\rho_{b}$) plus the resistivity $R_{t}$ and the photoelectric factor $pef$ or the sonic $\Delta_{t}$. Petrophysical models involve usually between 10 and 15 components. We order them in different classes of minerals: Shale, Sand, Carbonate. The porosity $\phi$, containing the different fluids (water, oil or gas), is added to the model.  This classification is important for the prior of the lithological model. Table \ref{table:litho_model} illustrates an example of a lithological model with the main families of minerals.
\begin{table}[H]
 \fontsize{8}{12}\selectfont

\begin{tabu}{|X|X|}
\hline Family & Components\\
	 \hline\hline

		Sand-Mica &  Quartz, Plagioclase, Mica, Feldspar \\ 
		\hline                
		Carbonate & Calcite, Dolomite, Ankerite\\ 
		\hline
		Clay/Shale  & Illite, Chlorite, Smectite, Kaolinite...\\ 
		\hline
		Porosity & Water, Oil or Gas  \\ 
		\hline
		Others & Halite, Anhydrite, Pyrite.. \\ 

	 \hline\hline
\end{tabu}
\caption{Example of lithological model}
\label{table:litho_model}
\end{table}

\FloatBarrier
\subsection{Method}
\begin{figure}[!h]
	\centering
		\includegraphics[width=\textwidth]{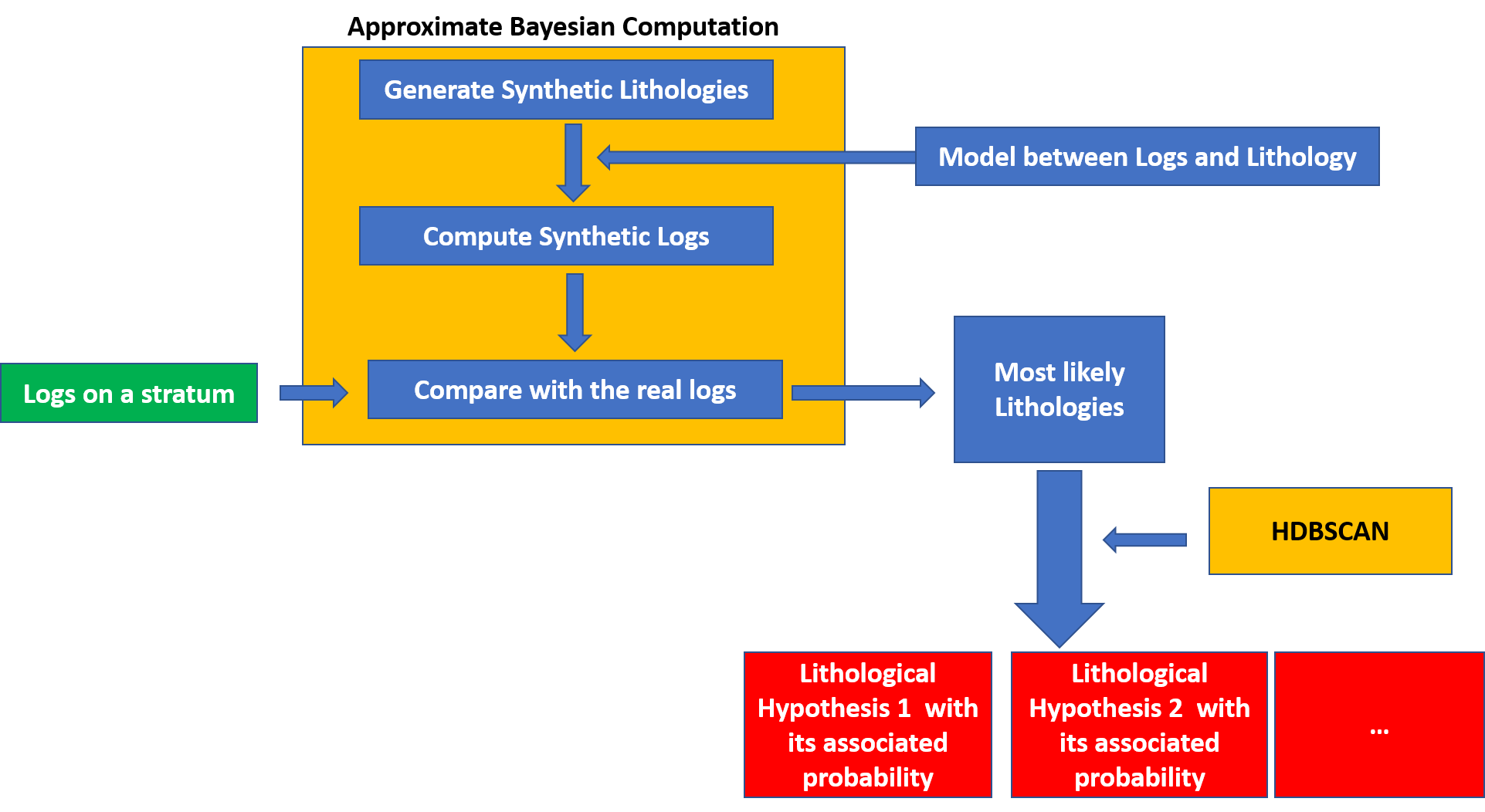}
	\caption[Global methodology for mineral inversion]{Global methodology: from the logs of a layer to the petrophysical hypothesis. The inputs, the methods used and the outputs are respectively in green, yellow and red. }
	\label{fig:global_method}
\end{figure}

We propose in this section a two-step method based on Approximate Bayesian Computation or ABC (see \cite{abc} for a review of this Bayesian method) and density-based clustering to get the lithological hypothesis on a given stratum. Figure \ref{fig:global_method} summarizes the proposed methodology.
\FloatBarrier

\subsubsection{Bayesian inference with ABC} \label{method}
\paragraph{About the Dirichlet distribution}
The Dirichlet distribution of order $k$, $\operatorname{Dir}(\alpha)$ where $\alpha \in \mathbb{R}^{k}$, allows us to generate a vector $\textbf{x}=\left(x_1,\ldots,x_k\right) \in \mathbb{R}^{k} $ in the $k-1$ simplex, or in other words: $\sum_{i=1}^{k}x_i=1$ and  $ 0 \leq x_i$. Its probability density function is given by: 
\begin{equation*}
 f\left(x_1,\cdots,x_k; \alpha_1, \cdots, \alpha_k\right)\frac{1}{B(\alpha)}\prod_{i=1}^{M}x_i^{\alpha_i-1},
\end{equation*} 
where B is the beta function.  The expectation of each element of \textbf{x} is $\operatorname{E}[x_i] = \frac{\alpha_i}{\sum_{i=j}^{k}\alpha_j}$. The variance and the covariances are proportional to $ \frac{1}{\left(\sum_{j=1}^{k}\alpha_j\right)^{2}}$.

\begin{figure}[!h]
   \begin{minipage}[c]{.45\linewidth}
		\includegraphics[width=\textwidth]{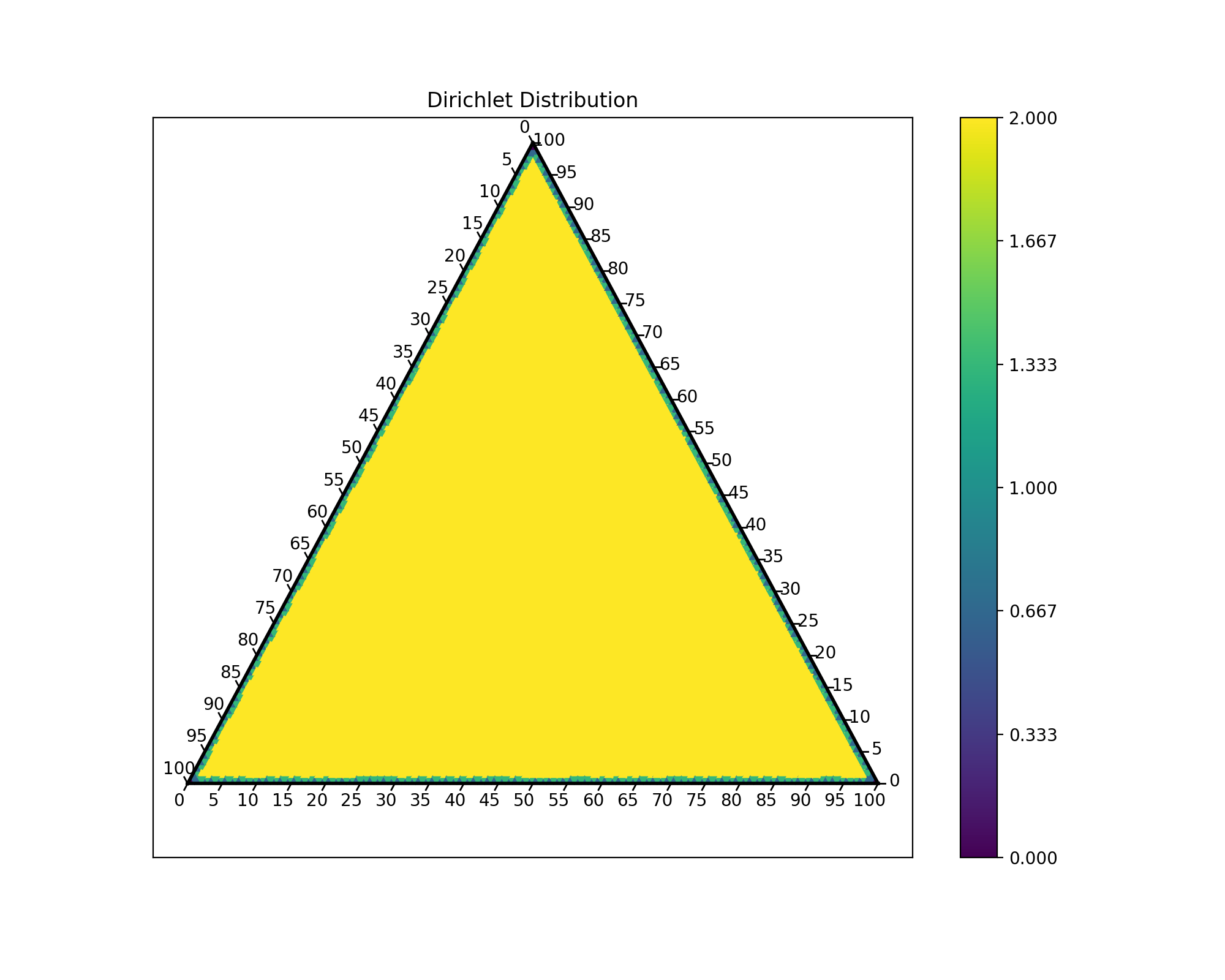}
	\caption[Ternary Density plot of a Dirichlet distribution]{Ternary Density plot of a $\operatorname{Dir}(1,1,1)$. It is a uniform distribution
over the simplex of $\mathbb{R}^{3}$.}
		\label{fig:dirichlet_ditribution_1}
   \end{minipage} \hfill
   \begin{minipage}[c]{.45\linewidth}
		\includegraphics[width=\textwidth]{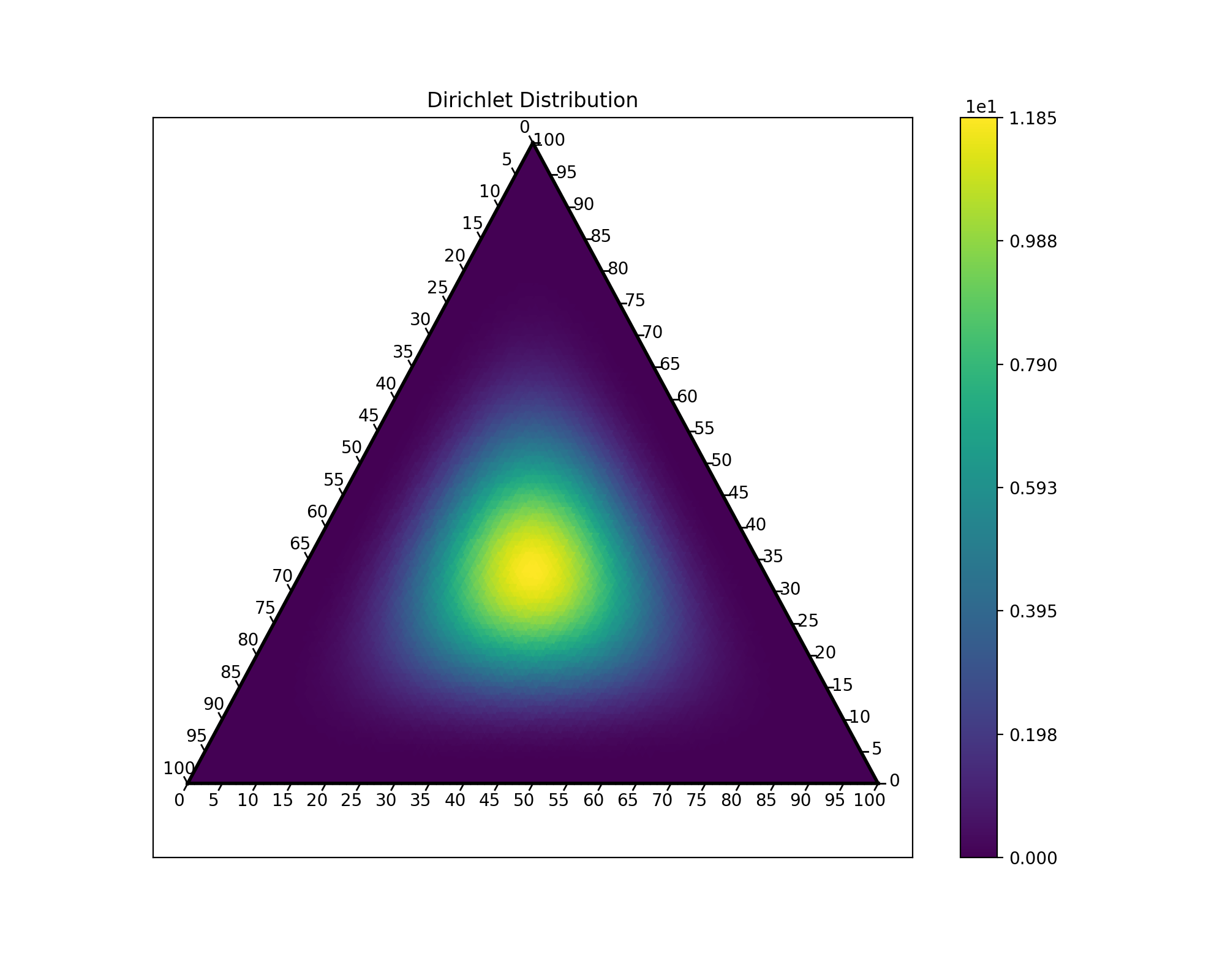}
		\caption[Ternary Density plot of a Dirichlet distribution]{Ternary Density plot of a $\operatorname{Dir}(5,5,5)$. We observe a mode at $\left(\frac{1}{3},\frac{1}{3},\frac{1}{3}\right)$.}
			\label{fig:dirichlet_ditribution_5}
   \end{minipage}
\end{figure}

When $\alpha_1 = \ldots=\alpha_k = 1$, $\operatorname{Dir}(\alpha)$ is the uniform distribution on the  $k-1$ simplex. When $\alpha_1 = \ldots = \alpha_k = 0.1$, sparsity appears over the simplex, the corners and the edges of the simplex have more density mass. If the $\alpha_i > 1$ ($\alpha_1 =\ldots = \alpha_k = 5$), a mode will appear clearly on the location of the average. Figures \ref{fig:dirichlet_ditribution_1} and \ref{fig:dirichlet_ditribution_5} illustrate these cases in $\mathbb{R}^{3}$.
\FloatBarrier
 
The inference model is based on a Dirichlet distribution as prior for the mineralogy ($V$) and introduces a Gaussian noise for the logs ($L$).  The Dirichlet distribution fulfills  the closure constraints of  the volumes. The model can be specified as : 
\begin{equation}
  \left\{
      \begin{aligned}
       V &\sim \mathcal{D}\left( \alpha_{1},\ldots \alpha_{M}\right), \\
       L &\sim \mathcal{N}\left(G\left(V\right) , \Sigma\right).\\
      \end{aligned}
    \right.
\end{equation}
Once for all, we set the known operator G as linear and bounded, so $G  \in \mathcal{M}_{d,M} \left( \mathbb{R} \right)$ the set of rectangle real-valued matrices with $d$ rows and $M$ columns. The elements of $G$ are the cells $\mathcal{M}_{d,M}$ and stand for the endpoints.
We can rewrite this model in terms of likelihood function:
\begin{equation}
  \left\{
      \begin{aligned}
       p\left(V\right) &= \frac{1}{B(\alpha)}\prod_{i=1}^{M}V_i^{\alpha_i-1}, \\
       p\left(L\vert V\right) &= \exp\left[ -\frac{1}{2}\left( GV -L\right)^{T}\Sigma^{-1}\left(GV-L\right)\right].\\
      \end{aligned}
    \right.
\end{equation}
 Using Bayes' rule, the likelihood function of the combined model can be written as:
\begin{equation}
      \begin{aligned}
       p\left(V\right\vert L) &= \frac{p\left(L\vert V\right)p\left(V\right)}{p\left(L\right)},\\
       p\left(V\right\vert L) & \propto \exp\left[ -\frac{1}{2}\left( GV -L\right)^{T}\Sigma^{-1}\left(GV-L\right)\right]\frac{1}{B(\alpha)}\prod_{i=1}^{M}V_i^{\alpha_i-1}.\\
      \end{aligned}
\end{equation}
Because $d \ll M$, the matrix G is not invertible so we cannot compute the maximum a posteriori (MAP) estimation of the lithology V. To compute the  posterior distribution of the lithology we need a numerical method. Bayesian frameworks have been used to solve ill-posed inverse problem  because they handle the non-uniqueness of the solution (see \cite{stuart2010inverse}). In this study, we adopt the Approximate Bayesian Computation: it is a simple rejection method. Let $\mathbf{\theta}$ be the parameter of interest, $\mathbf{y} $  the observations, $\epsilon$ a tolerance level, $\pi$ the prior distribution of the model, and $\rho$ a distance function. The original algorithm of ABC is given below:
\begin{algorithm}[H]
  \caption{ General ABC}\label{abc_original}
  \begin{algorithmic}
	\For{$i=1$ to J}
	\Repeat
        \State \texttt{Generate $\theta^{\prime}$ from the prior distribution $\pi(.)$}
         \State \texttt{Generate $\mathbf{z}$ from the likelihood $f(.\vert\theta^{\prime} )$}

      \Until{$\rho\left(\mathbf{z},\mathbf{y}\right)\leq \epsilon$}
      \State \texttt{set $\theta_i=\theta^{\prime}$}
      \EndFor
        \end{algorithmic}
\end{algorithm}

For our problem, we adapt Algorithm \ref{abc_original}. At each depth $n$ we state :
\begin{algorithm}[H]
  \caption{ Adapted ABC}\label{abc}
  \begin{algorithmic}
	\For{$i=1$ to J}
        \State \texttt{Generate $V^{\prime}$ from the prior distribution }
         \State \texttt{Generate the corresponding $L^{\star}$ from the model}

      \If{$\forall 1 \leq i \leq d$, $\|L^{\star}_{i}-L_{n,i} \|<\delta_{i}$}
       \State \texttt{accept $V^{\prime}$}
      \EndIf
      \EndFor
        \end{algorithmic}
\end{algorithm}

Where the $\delta_{i}$ are defined for each log and should be calibrated and $J$ is around $10^{6}$.  
We apply the ABC procedure for all the depth of the layer and get a posterior distribution on the layer of the lithology. Due to the non-uniqueness of the inverse problem, the posterior probability on the layer of lithology is often multi-modal. For instance, if  two shales similar in terms of physical parameters appear in the lithological model,  it will be hard to differentiate them.

\subsubsection{Clustering on the results of ABC} \label{clustering}
In order to retrieve the most probable lithology hypothesis on the given stratum, we perform a density-based clustering algorithm on the ABC results. The aim of the clustering is to distinguish the modes that may appear in the results of ABC. Ideally we seek a density-based clustering algorithm detecting the outlying lithologies and with a data-driven tuning of the number of clusters.  In the literature, several generic density-based algorithms may be found such as density-based spatial clustering of applications with noise (DBSCAN) \cite{dbscan}, ordering points to identify the clustering structure (OPTICS)  \cite{optics} and  Hierarchical DBSCAN or HDBSCAN \cite{hdbscan}. We choose HDBSCAN for many reasons:
\begin{itemize}
\item Conversely to DBSCAN, the distance value (main parameter of DBSCAN) disappears. The main parameter is the minimum size of the cluster.
\item Identification of the outliers or noise is available in DBSCAN but not in OPTICS.
\item The \textit{hdbscan} package in Python (\cite{mcinnes2017hdbscan}) is quite efficient, could be run in parallel and offers low computational time. The size of a stratum could reach sometimes hundreds of feet so we may have  100 000 points to cluster.
\end{itemize}
The two main parameters of HDBSCAN are: the number of points $m_{pts}$ used for the core
distance (DBSCAN, OPTICS) and the minimum size to form a cluster $m_{clSize}$. Usually,
$m_{clSize} = m_{pts}$ and having a smaller $m_{pts}$ implies fewer points detected as noise. Here is a description of the algorithm:
\begin{algorithm}[H]
  \caption{HDBSCAN}\label{hdbscan_algo}
  \begin{algorithmic}
	        \State \texttt{Compute $ \forall x_{p}$ the core distance $d_{core,m_{pts}}(x_{p})$}
         \State \texttt{Build a minimum spanning tree (MST) using the  mutual reachability distance  \\$d_{mreach,m_{pts}}(x_{p},x_{q})$}
					\State \texttt{Derive the cluster hierarchy from the MST. Transform the MST into a dendrogram}
     \State \texttt{Condense the dendrogram using $m_{clSize}$}
        \end{algorithmic}
\end{algorithm}

\begin{figure}[!h]
	\centering
		\includegraphics[scale=0.7]{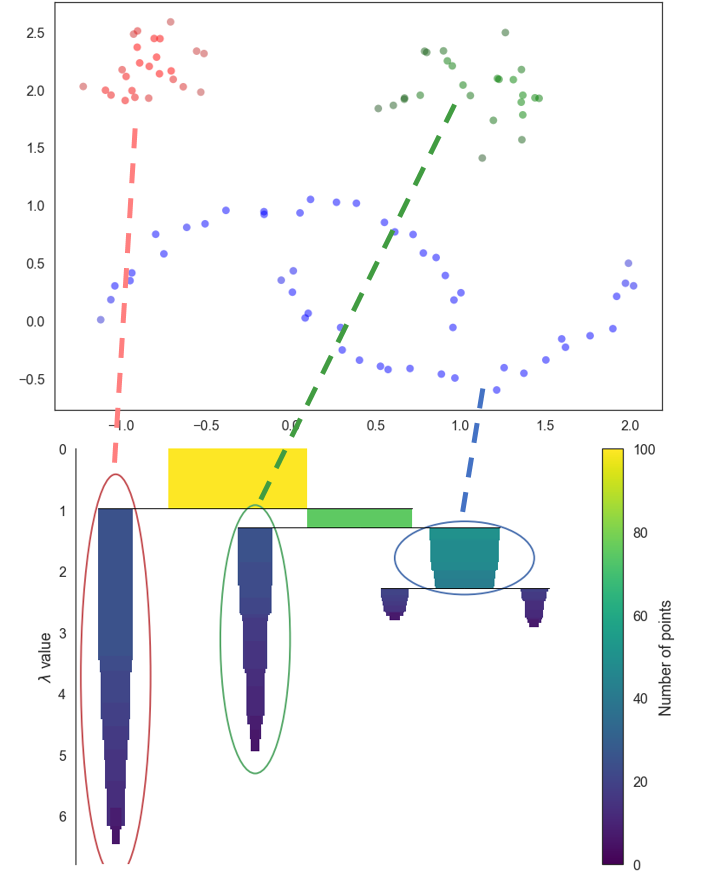}
	\caption[Example of condensed tree on the two moons dataset]{Example of a condensed tree on the two moons dataset (toy example of the \textit{hdbscan} library in python \cite{mcinnes2017hdbscan}). We have the three coloured clusters found by HDBSCAN. The width of different nodes and leaves represent the number of elements. The encircled nodes and leaves are the clusters selected by the EOM criterion. The $\lambda$ axis represents the density of the cluster.}
	\label{fig:condensed_tree}
\end{figure}

The core distance of a point $x_p$, $d_{core}(x_{p})$, is the distance from $x_{p}$ to its $m_{pts}$-nearest neighbor (including $x_{p}$). The mutual reachability distance between two points $x_{p}$ and $x_{q}$ in $\bold{X}$, with regards to $m_{pts}$, is defined as $d_{mreach}(x_{p}, x_{q}) =\max \lbrace d_{core}(x_{p}), d_{core}(x_{q}), d(x_{p}, x_{q})\rbrace$. An output of HDBSCAN is the condensed tree which gives a hierarchical visualization of the data in form of a tree. The Excess of Mass (EOM) criterion described in \cite{hdbscan_2015} allows a flat clustering by selecting the stable leaves of the condensed tree as clusters.  Figure \ref{fig:condensed_tree} is  an example of condensed tree. The height of the leaves represents the local density.

With HDBSCAN,  defining the minimum size of a cluster is mandatory, we choose $m_{clSize}=m_{pts}$ between 1 and 5$\%$ of the number of points. This parameter has to be calibrated. We can order the different clusters of lithology by their size and/or the error of reconstruction of a classical inversion method when the number of lithological component  obeys $M \leq d+1 $. 

We can define an empirical probability $\hat{p}_{i}$ for each lithological hypothesis:
\begin{equation}
 \hat{p}_{i}=\frac{m_{i}}{N_{ABC}},
\end{equation}
where $N_{ABC}$ is the number of lithology selected by the ABC part and $m_{i}$ is the size of the cluster i found by HDBSCAN. Note that $\sum \hat{p}_{i}< 1$ because HDBSCAN considers a part of the lithologies as noise.

\section{Results on synthetic data} \label{synth.data}

\subsection{Synthetic example and  Monte Carlo sampler}

 We use below a synthetic generator of logs that may be roughly summarized in three steps. First, we generate an average lithology using a Dirichlet distribution, and then we add a Brownian bridge to get a synthetic lithology over the synthetic stratum. Finally, we transform the lithology into logs using equation \ref{eq:log_equation} and we add noise.
  
\begin{figure}[!h]
	\centering
		\includegraphics[scale=0.6]{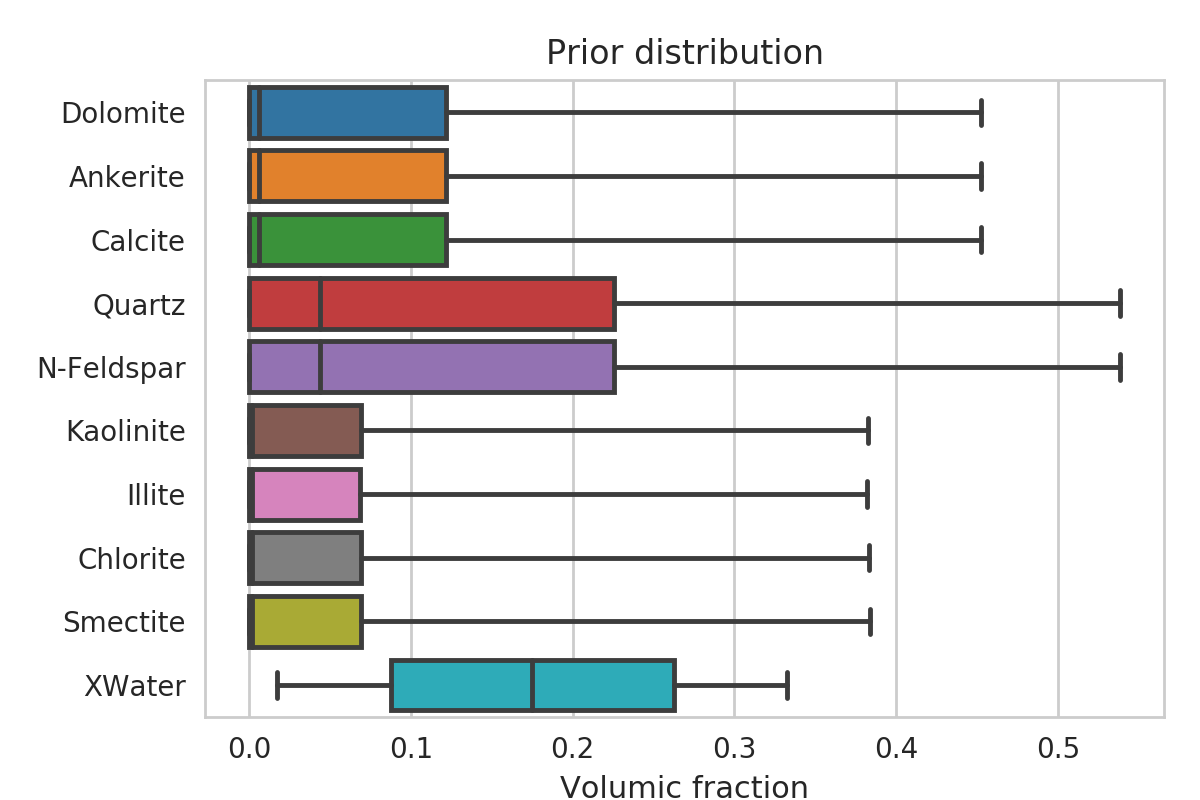}
	\caption[Example of prior distribution of the lithology]{Example of prior distribution of the lithology. On the boxplots, we represent the percentile 5 and percentile 95 as the whiskers (applied to all the box plots).}
	\label{fig:prior_distribution}
\end{figure}

In our test with ABC, we select $M=10$ lithological components: 3 carbonates (Calcite, Ankerite, Dolomite), 2 sands (Quartz, N-Feldspar), 4 shales (Illite, Kaolinite, Chlorite, Smectite) and water for the porosity. Here we do not use a  simple Dirichlet distribution where all the $\alpha_{i}$ are equal. Indeed, the porosity is generally below 35$\%$ and we assume that a mix of minerals of the same family  is not probable. For this reason, we draw first $V_{water}\sim \mathcal{U} \left[ 0,0.35 \right]$, then the proportion of the families (carbonate, shale and sand) $\left(V_{sand}, V_{shale}, V_{carbonate}\right)\sim  \left(1-V_{water}\right) \mathcal{D}\left( 1,1,1\right)$. Finally, we draw the minerals according to a Dirichlet distribution with a small alpha parameter equal to 0.1. For instance, we generate the volumes of the carbonate family: $\left(V_{calcite},V_{dolomite},V_{ankerite}\right) \sim V_{carbonate} \mathcal{D}\left( 0.1,0.1,0.1\right)$. Indeed, we prefer a sparse model fitting the reality: sparsity is a well-known feature of most lithological databases. We defined a lithological sampling model. An example of prior distribution on the lithology using the sampling model is shown in Figure \ref{fig:prior_distribution}.  Besides, we generate 1 million lithologies for ABC.

\begin{table}[H]
 \fontsize{8}{12}\selectfont

\begin{tabu}{|X[3.0cm]|X|X|X|X|X|X|X|X|X|}
\hline Volume &   Chlorite  &Illite & Kaolinite & Smectite & Quartz  & Calcite & Dolomite  & XWater& XOil\\
	 \hline

		Sandy & - & -  &- & - & 80 & - & - & 20 &-   \\                 
		Sandy Oil & - & -  &- & - & 80 & - & - & 5 &15   \\ 
		Shaly-Sand 1 & - & 40  &- & - & 40 & - & - & 20&-   \\ 
		Shaly-Sand 2 & - & 80  &- & - & 10 & - & - & 10 &-  \\ 
		Shaly-Sand 3 & - & -  &- & 80& 10 & - & - & 10  &- \\ 
		Shaly-Sand 4 & 40 & 40  &- & -& 10 & - & - & 10  &- \\ 

		Shaly-Carbonate 1  & - & -  &- & 60& - & 20 & - & 20&-   \\ 
		Shaly-Carbonate 2 & - & -  &30 & - & - & 50 & - & 20  &- \\ 
		Shaly-Carbonate 3 & - & 20  &20 & -& - & - & 40 & 20 &-  \\ 
		Carbonate-Shaly  & - & 20  &- & - & - & 30 & 30 & 20 &-  \\ 
		Carbonate  & - & -  &- & - & - & 40 & 40 & 20&-   \\ 
	\hline
\end{tabu}
\caption[Different lithologies tested]{Different lithologies tested. The lines correspond to the alpha parameter. They are similar to the proportion. }	
\label{table:model}
\end{table}

\begin{figure}[!h]
   \begin{minipage}[c]{.45\linewidth}
		\includegraphics[width=\textwidth]{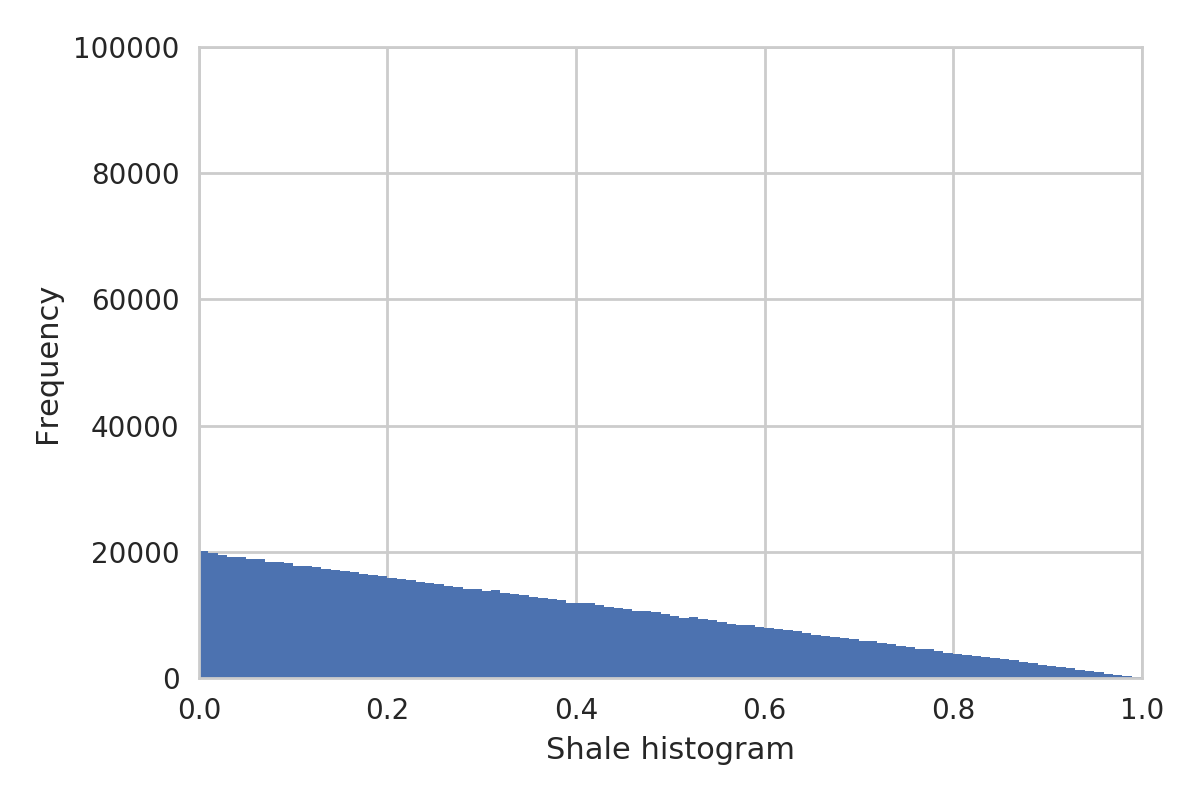}
	\caption[Histogram of the shale volume]{Histogram of the shale volume generated by the sampling model.}
		\label{fig:shale_hist}
   \end{minipage} \hfill
   \begin{minipage}[c]{.45\linewidth}
		\includegraphics[width=\textwidth]{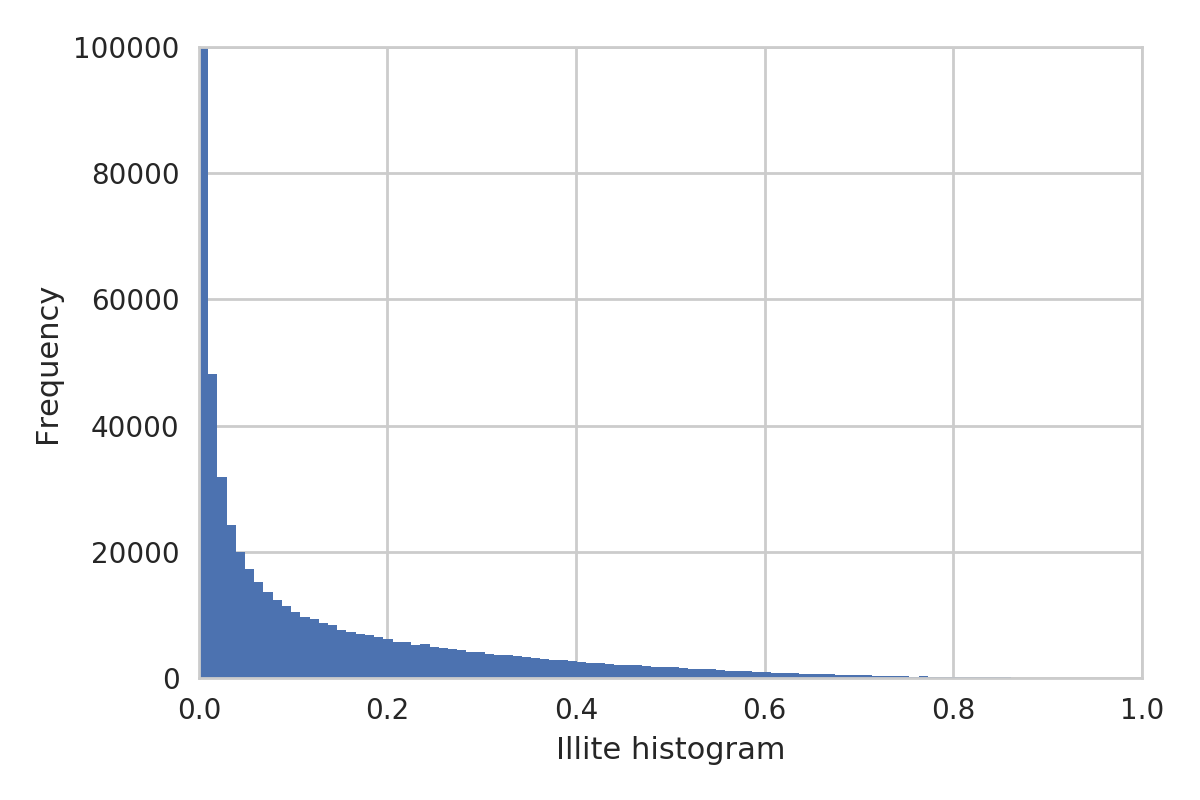}
		\caption[Histogram of the Illite volume]{Histogram of the Illite volume generated by the sampling model. The frequency around 0 is near 600 000 samples.}
			\label{fig:illite_hist}
   \end{minipage}
\end{figure}

Here are some examples of distributions that the defined sampling model generates over 1 million samples. The histogram of the total volume of shale is presented in Figure \ref{fig:shale_hist}, the probability near one is very low because of the water distribution which is not correlated to the mineral volumes. Figure \ref{fig:illite_hist} displays the histogram  of the volume of illite generated by the sampling model. An explanation of the sampler parameter is that an important variance for the volume fits well the reality. Indeed, according to the Dirichlet distribution with $\alpha_{1}=\cdots=\alpha_{10}=1$, $\mathrm{Var}\left[V_{i}\right]=0.008185$ and if $\alpha_{1}=\cdots=\alpha_{10}=0.1$ then $\mathrm{Var}\left[V_{i}\right]=0.045$. With the defined sampling model  $\mathrm{Var}\left[V_{i}\right]\simeq0.02$ (depending on the family of the mineral) except for the water $\mathrm{Var}\left[V_{XWater}\right]=0.01$).
\FloatBarrier

\subsection{Results}
\begin{figure}[!h]
	\centering
		\includegraphics[scale=0.6]{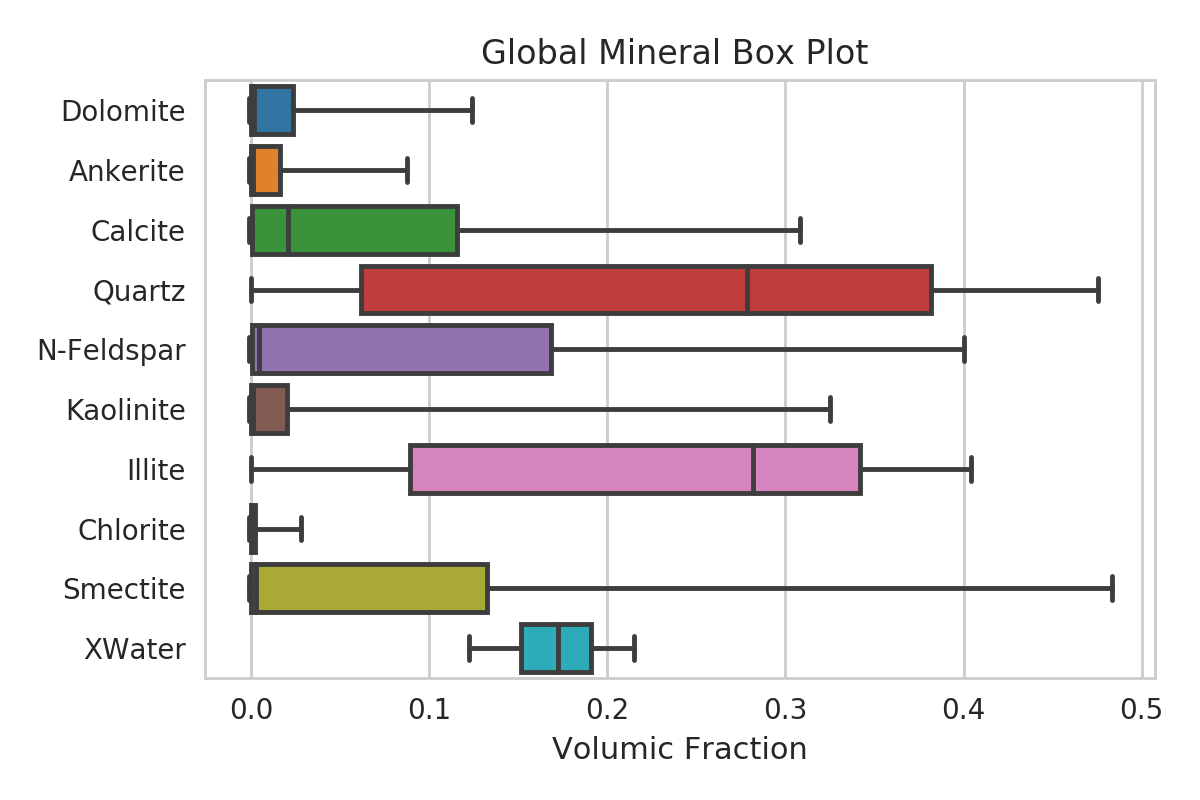}
	\caption{Boxplot of the volumes selected on the stratum by ABC. }
	\label{fig:boxplot_global}
\end{figure}

\begin{figure}[!h]
  \centering
   \begin{minipage}[t]{.45\linewidth}
		\includegraphics[width=\textwidth]{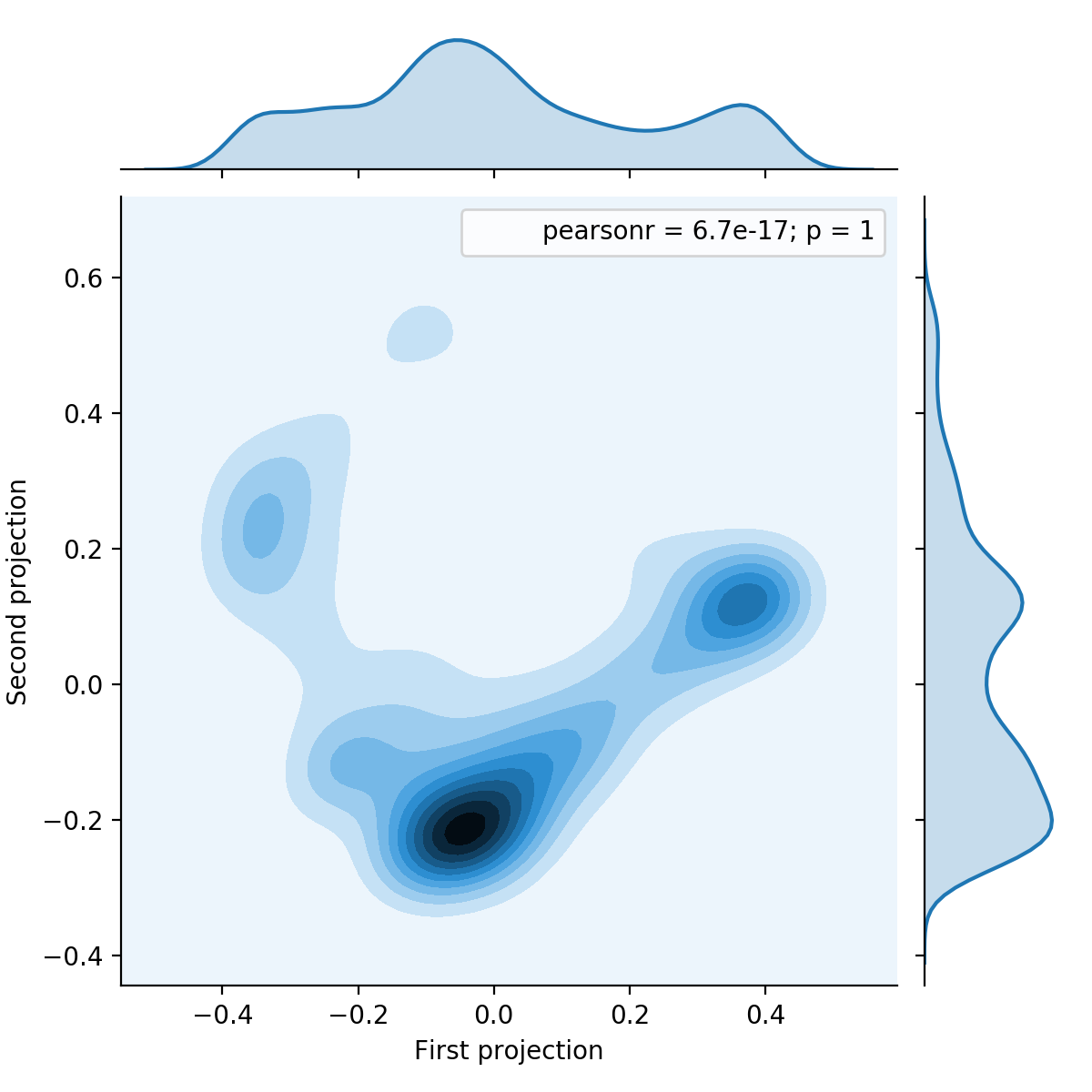}
		\caption{Density cross-plot of the first and second component of the PCA. }
			\label{fig:pca_density}
		
   \end{minipage}\hfill
    \begin{minipage}[t]{.45\linewidth}
    \includegraphics[width=\textwidth]{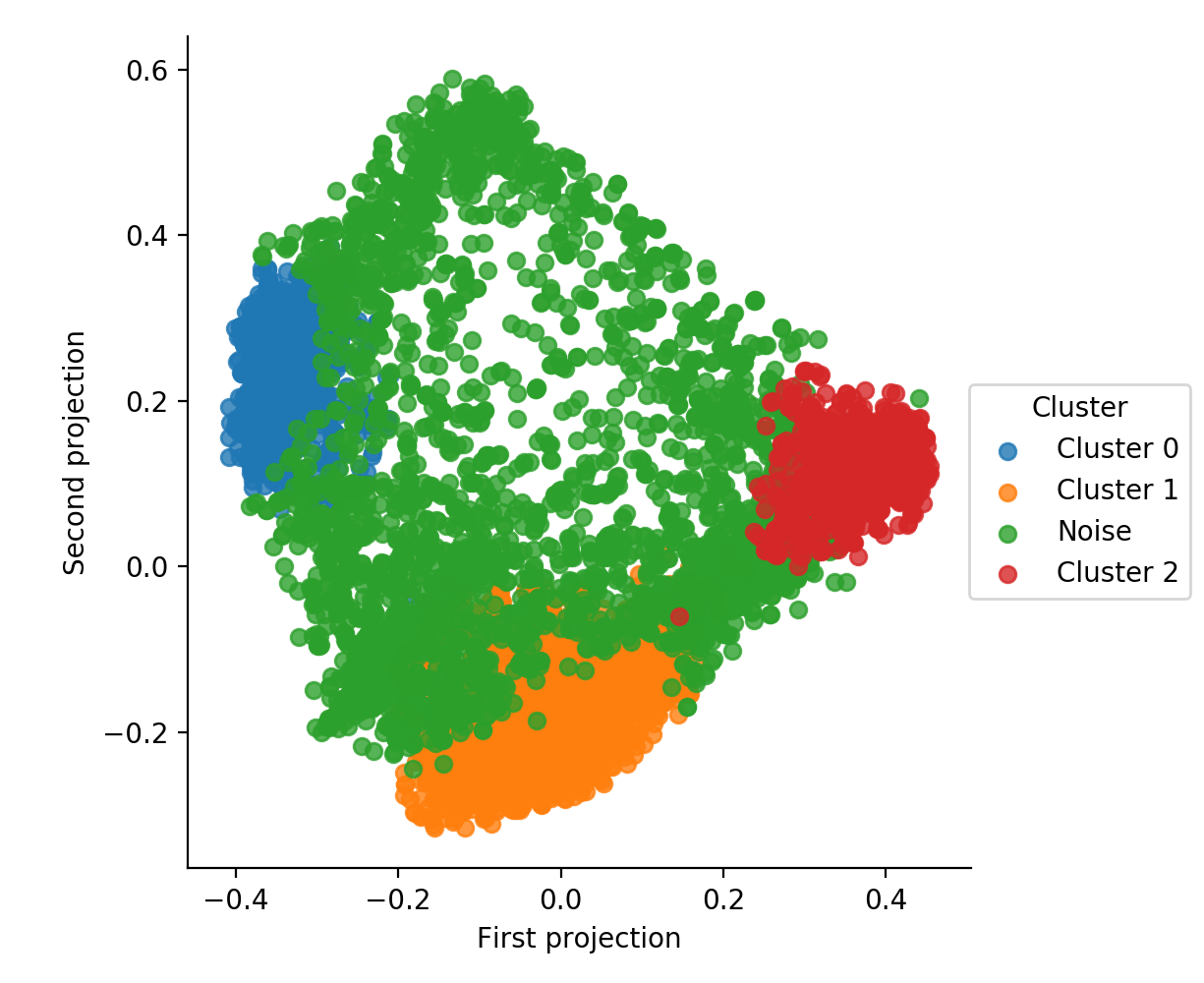}
	\caption{Cluster found by HDBSCAN (PCA projection).}
		\label{fig:hdbscan_cluster}
    
   \end{minipage} 
\end{figure}

We show here the results on the Shaly-Sand 1 (see Table \ref{table:model}) example with a stratum of 250 samples and with the following logs: gamma ray GR, bulk density $\rho_{b}$, neutron porosity $\phi_{N}$. On average we have $V_{Illite}=0.34$, $V_{Quartz}=0.49$ and $V_{Water}=0.17$.
With a fixed rejection factor $\delta$ on the logs ($\delta_{GR}=12$ API, $\delta_{\rho_b}=0.05$ $G/C^{3}$, $\delta_{\phi_{N}}=0.03$ V/V), we obtain an average of 3000 lithologies per depth. In Figure \ref{fig:boxplot_global}, the boxplot of all the volumes selected by ABC on the stratum is provided.  We see that the volumes of Illite, Smectite, Quartz, and N-Feldspar have a large variance. We remark too that the volumes of carbonate are low. For more details on these distributions, we look at the histograms of the first and second components of the principal component analysis (PCA) of the results. The percentage of variance explained by the first and second components are respectively 40\% and 33\%. Figure \ref{fig:pca_density} displays a cross-plot of the two components and the associated  histograms. We notice the multi-modality of the first projection with a main mode and the  bimodality of the second projection. An explanation is that the modes of the first projection accounts for the competition between Kaolinite and Smectite. The second projection shows the competition between the Quartz and the N-Feldspar. 
After the ABC step, we apply HDBSCAN in order to find the most plausible lithological hypothesis. We fix the minimum number  of samples to form a cluster to 5\% of the number of the lithology selected by ABC. 
 In this case, HDBSCAN finds 3 clusters and classifies around 35\% lithologies as noise. Figures \ref{fig:hdbscan_cluster} illustrates the results of HDBSCAN on the first and second component of the PCA of the volumes selected previously. We display the clustering on the cross-plot between the first and the second projections of the PCA. We distinguish clearly the clusters. 

\begin{figure}[!h]
   \begin{minipage}[c]{.45\linewidth}
		\includegraphics[width=\textwidth]{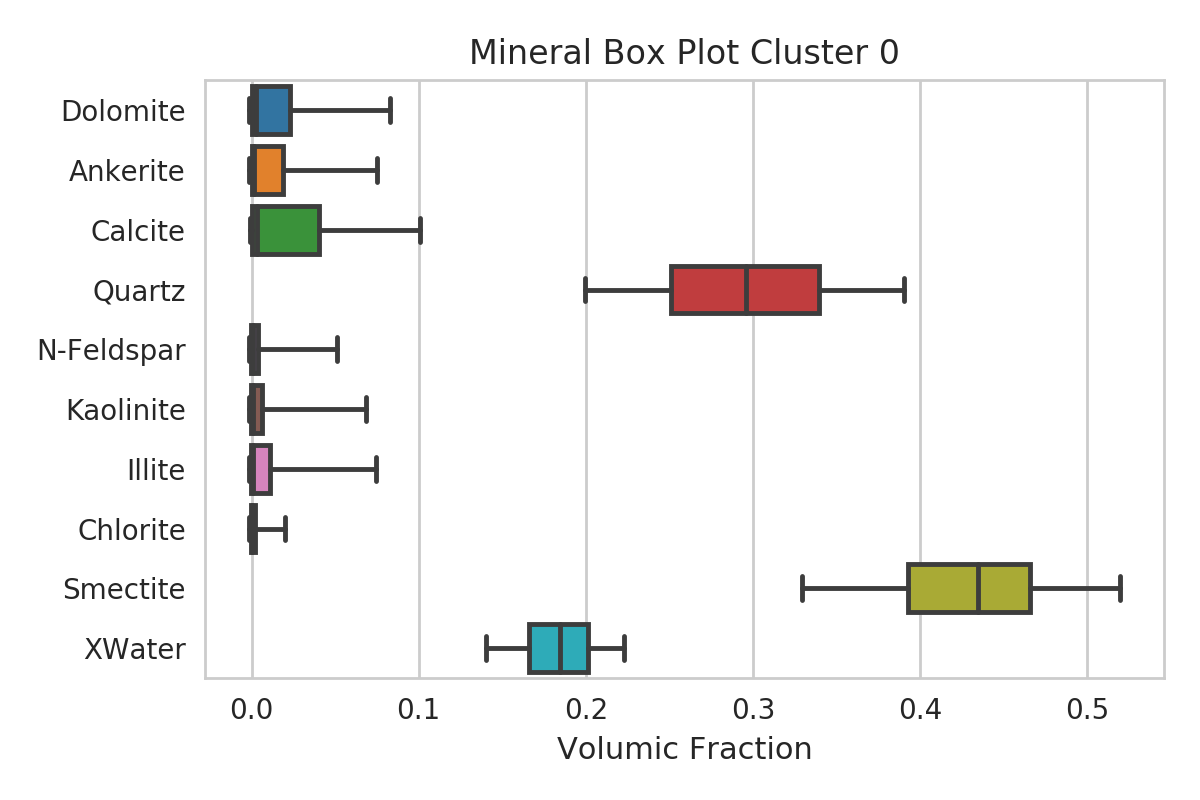}
	\caption{Boxplot of the volumes of cluster 0 (HDBSCAN).}
		\label{box-plot-cluster0}
   \end{minipage} \hfill
   \begin{minipage}[c]{.45\linewidth}
		\includegraphics[width=\textwidth]{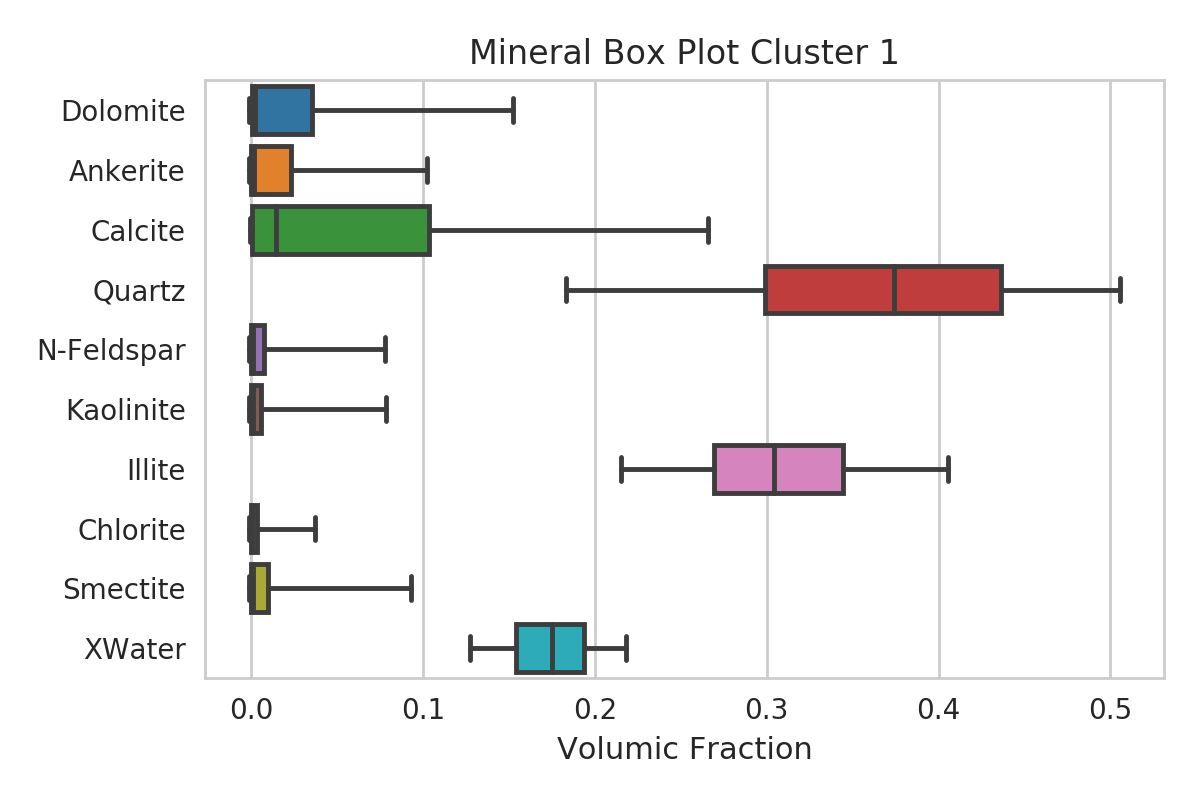}
		\caption[Boxplot of the volumes of the cluster 1 (HDBSCAN)]{Boxplot of the volumes of cluster 1 (HDBSCAN), the main cluster.}
			\label{box-plot-cluster1}
   \end{minipage}
   \hfill
   \begin{minipage}[c]{.45\linewidth}
		\includegraphics[width=\textwidth]{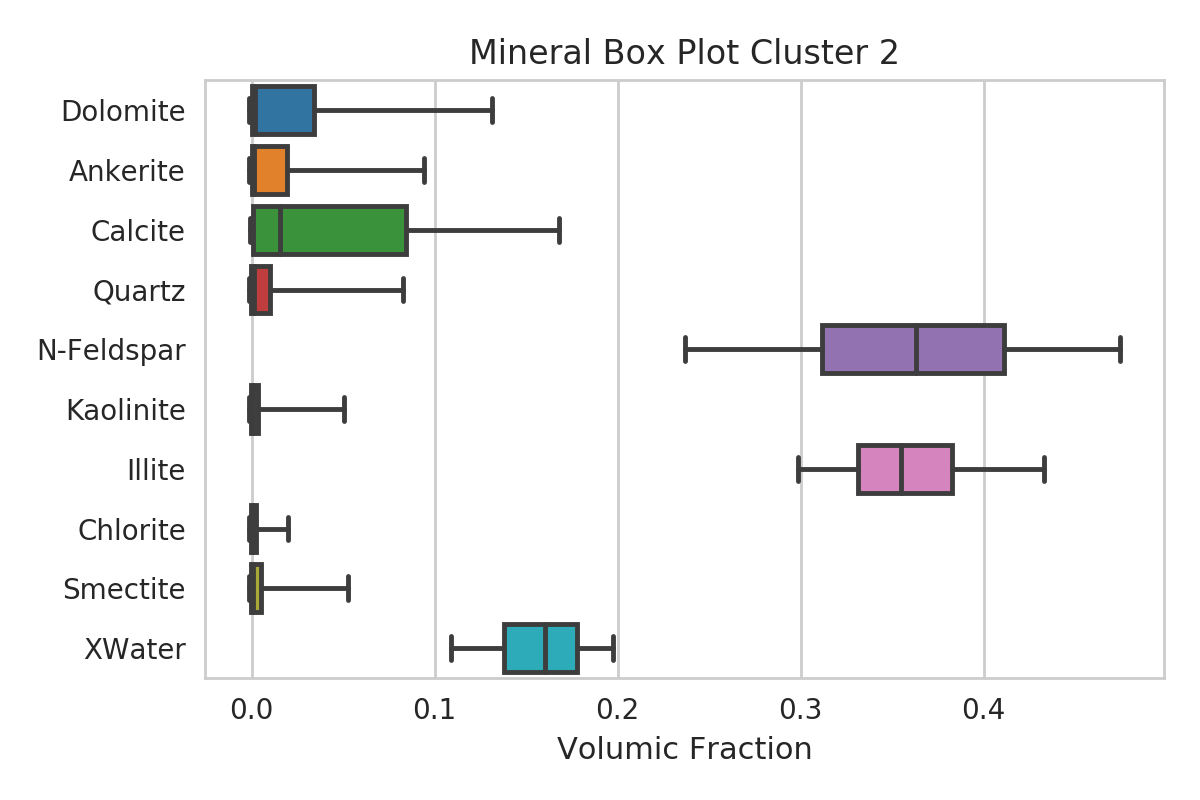}
		\caption{Boxplot of the volumes of cluster 2 (HDBSCAN).}
			\label{box-plot-cluster2}
   \end{minipage}
\end{figure}

 The boxplots of the three clusters are shown in Figures \ref{box-plot-cluster0}, \ref{box-plot-cluster1} and \ref{box-plot-cluster2}. Figure \ref{box-plot-cluster1} is the boxplot of the cluster with some Illite and Quartz corresponding to the synthetic lithology. In this cluster, the average of the three main components are: $\bar{V}_{Illite}=0.30$, $\bar{V}_{Quartz}=0.37$ and $\bar{V}_{XWater}=0.17$. The presence of Calcite, with an average volume $\bar{V}_{Calcite}=0.05$ explains why we underestimate the real volume of Quartz and Illite. A possible reason for this fact relies on the flat clustering of HDBSCAN. It tends to advantage large clusters and as consequence, we still have some competitions between Carbonate and Quartz which may have a similar response to the logs.  Cluster 0 (Figure \ref{box-plot-cluster0}) and Cluster 2 (Figure \ref{box-plot-cluster2}) are other possible lithological hypotheses where respectively  Illite is replaced by another shale, Smectite, and Quartz is replaced by another sand, N-Feldspar.

\begin{figure}[!h]
	\centering
		\includegraphics[scale=0.6]{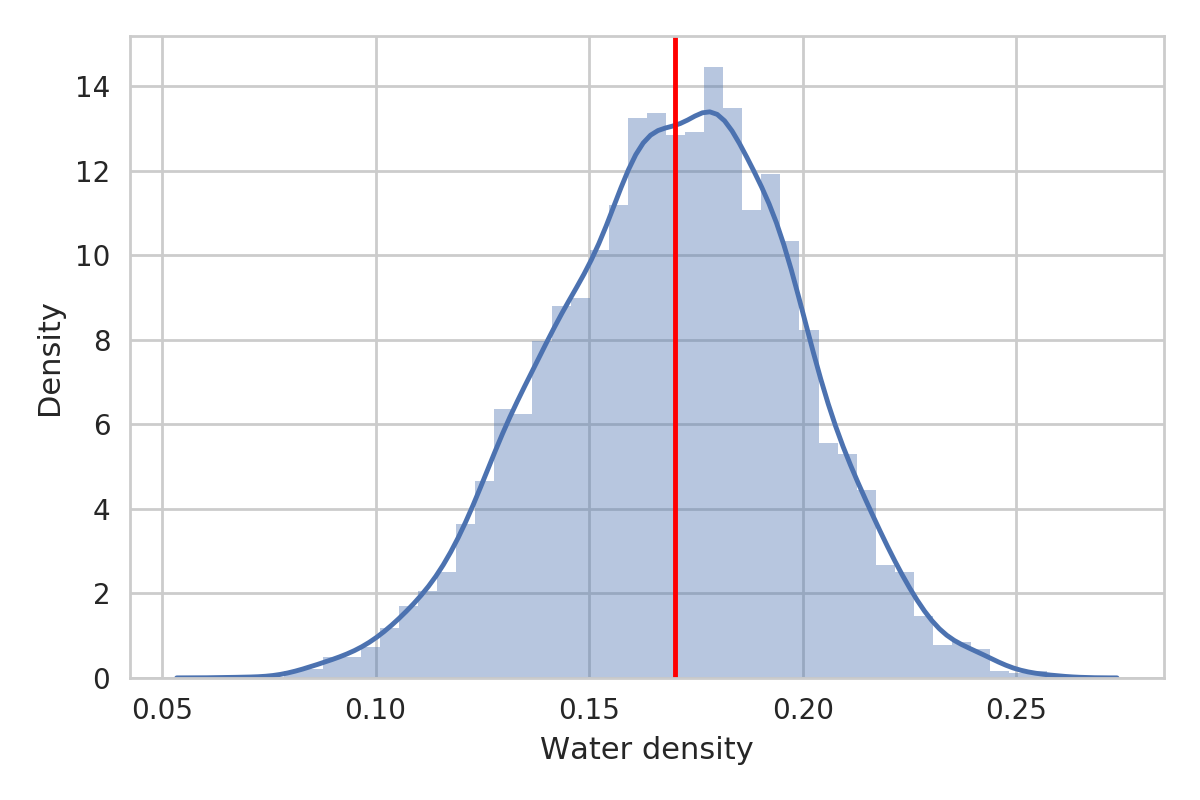}
	\caption[ Empirical distribution of the water volume]{ Empirical distribution of the water volume. We can derive from this plot an optimistic, average and pessimistic scenarii (percentile 10, median and percentile 90). The true average volume of water is 0.17 (vertical line in red). }
	\label{fig:density_water}
\end{figure}

After removing the 35 \% of lithology considered as noise, we can order the different hypotheses by the number of points:
\begin{itemize}
  \item Cluster 1 ($\hat{p}_1=40\%$). Main minerals: Quartz/Illite/Water.
  \item Cluster 2 ($\hat{p}_2=15\%$). Main minerals: N-Feldspar/Illite/Water.
  \item Cluster 0 ($\hat{p}_0=10\%$). Main minerals: Quartz/Smectite/Water.
\end{itemize}
Here the classification by the size of the hypothesis gives a good result: Cluster 1 is the lithological hypothesis (Quartz/Illite/Water) matching the real lithology. Our methodology can provide either most probable lithological hypotheses or empirical distributions for the components common to all hypotheses. For instance, we can define the empirical distribution of the water $\hat{f}_{Water}$:
\begin{equation}
 \hat{f}_{Water}=\frac{\sum m_{i}\hat{f}_{i,Water}}{\sum m_{i}},
\end{equation}
where $\hat{f}_{i,Water}$ is the empirical distribution of the water in cluster $i$ and $m_{i}$ is the size for cluster $i$. Figure \ref{fig:density_water} shows the empirical distribution for our synthetic case. The mode of this distribution is closed to the real average value of the water of the synthetic data (red vertical line on the figure).

\begin{figure}[!h]
	\centering
		\includegraphics[width=\textwidth]{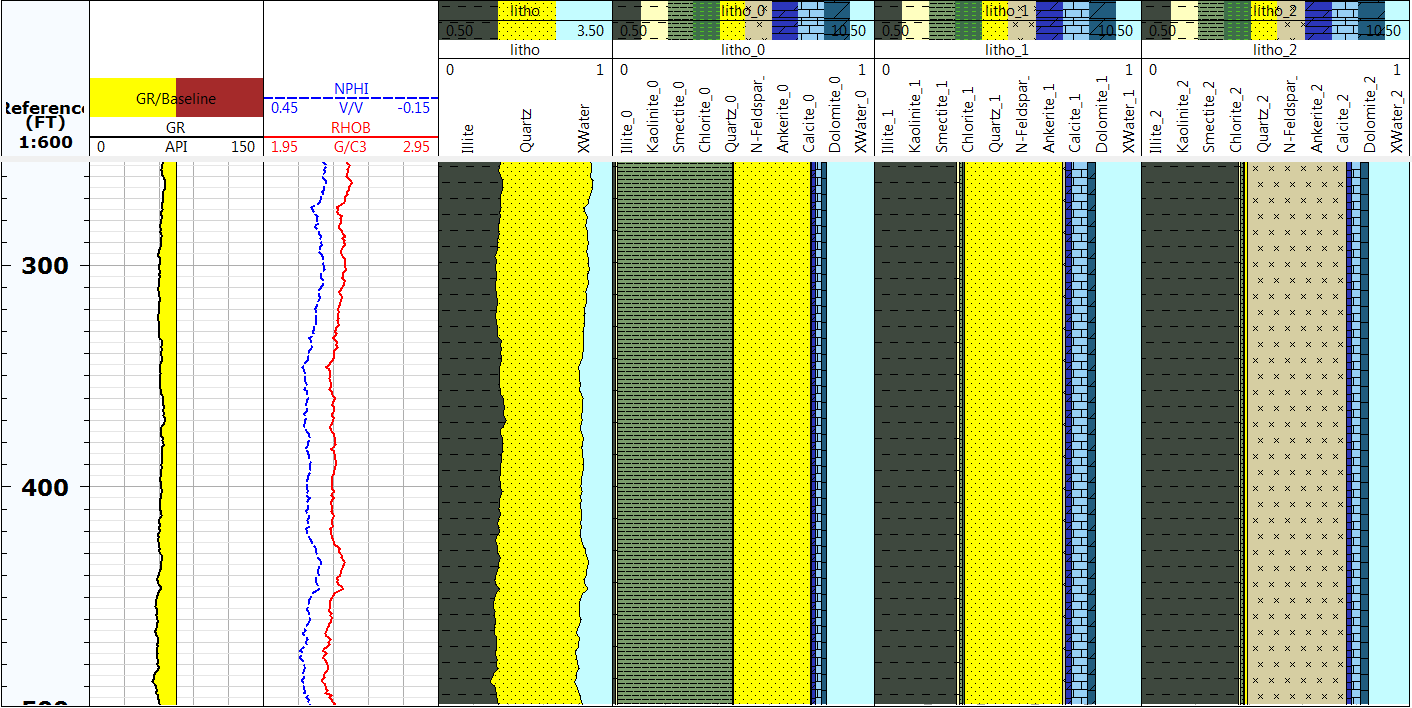}
	\caption[Synthetic lithology with its logs and results of our methodology]{Synthetic lithology with its logs and results of our methodology. Track 1: GR. Track 2: $\rho_{b}$ and $\phi_{N}$.   Track 3: Corresponding lithology. Track 4: Lithological hypothesis of cluster 0. Track 5: Lithological hypothesis of cluster 1. Track 6: Lithological hypothesis of cluster 2.}
	\label{fig:final_logiew}
\end{figure}

In Figure \ref{fig:final_logiew}, we present the synthetic lithology and the corresponding logs used to perform a Bayesian inversion per layer. In the last three tracks, the different hypotheses found by HDBSCAN (average of the minerals per hypothesis) are given.

For the other synthetic cases, corresponding to 2 minerals of different families and water, we are able to determine the right lithology. The volumes found are close to the reality.  For the mix of carbonate, sand or shale, the true hypothesis is not necessary the main one (it depends on the choice of the $\delta$ and the $\alpha$ parameter). Changing the parameter alpha has some consequences on the results. Moreover, when 3-4 components (shaly-sand-carbonate) are detected and one of them is around 10$\%$ it seems difficult to identify the smallest one.

\section{Case study } \label{case.study}
The dataset presented in this section comes from the Kansas Geological Survey, Oil and Gas Well Database (\url{http://www.kgs.ku.edu/Magellan/Qualified/index.html}). We present here a well from the Wellington Field: well 1-28 (API number: 15-191-22590). From this well, we get 9 logs: the gamma ray GR,  the resistivity $R_{t}$, the compressional slowness $\Delta_{t}$, the bulk density $\rho_{b}$, the neutron porosity $\phi_{N}$, the photoelectric factor $pef$, the uranium concentration $U$, the thorium concentration $TH$	and the potassium concentration $K$. We have also a computed lithology and the mud log to check our results. These data are part of the South-central Kansas $C0_{2}$ project.

\begin{figure}[!h]
	\centering
		\includegraphics[scale=0.8]{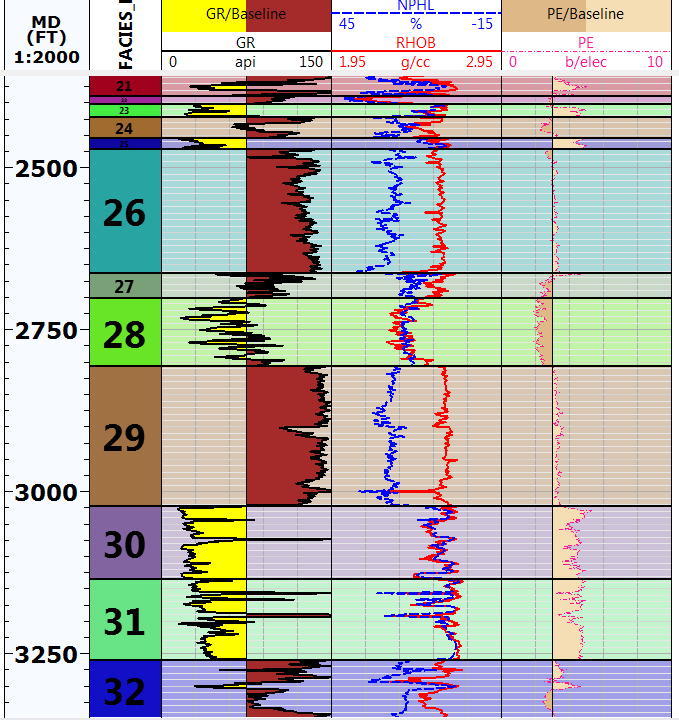}
	\caption[Well 1-28 with the zonation provided by PELT]{Well 1-28 with the zonation provided by PELT. Track 1: Zonation from PELT. Track 2: GR. Track 3: $\rho_{b}$ and $\phi_{N}$. Track 4: $pef$.}
	\label{fig:segmentation_1_28}
\end{figure}

For Well 1-28, we use the segmentation algorithm PELT (\cite{PELT}) to identify the different beds where we assume a fixed lithology. As inputs for PELT, we took GR, $\rho_{b}$, $\phi_{N}$ and $pef$ and a Gaussian model is used. Figure \ref{fig:segmentation_1_28} gives the results of the segmentation   

\begin{figure}[!h]
	\centering
		\includegraphics[scale=0.5]{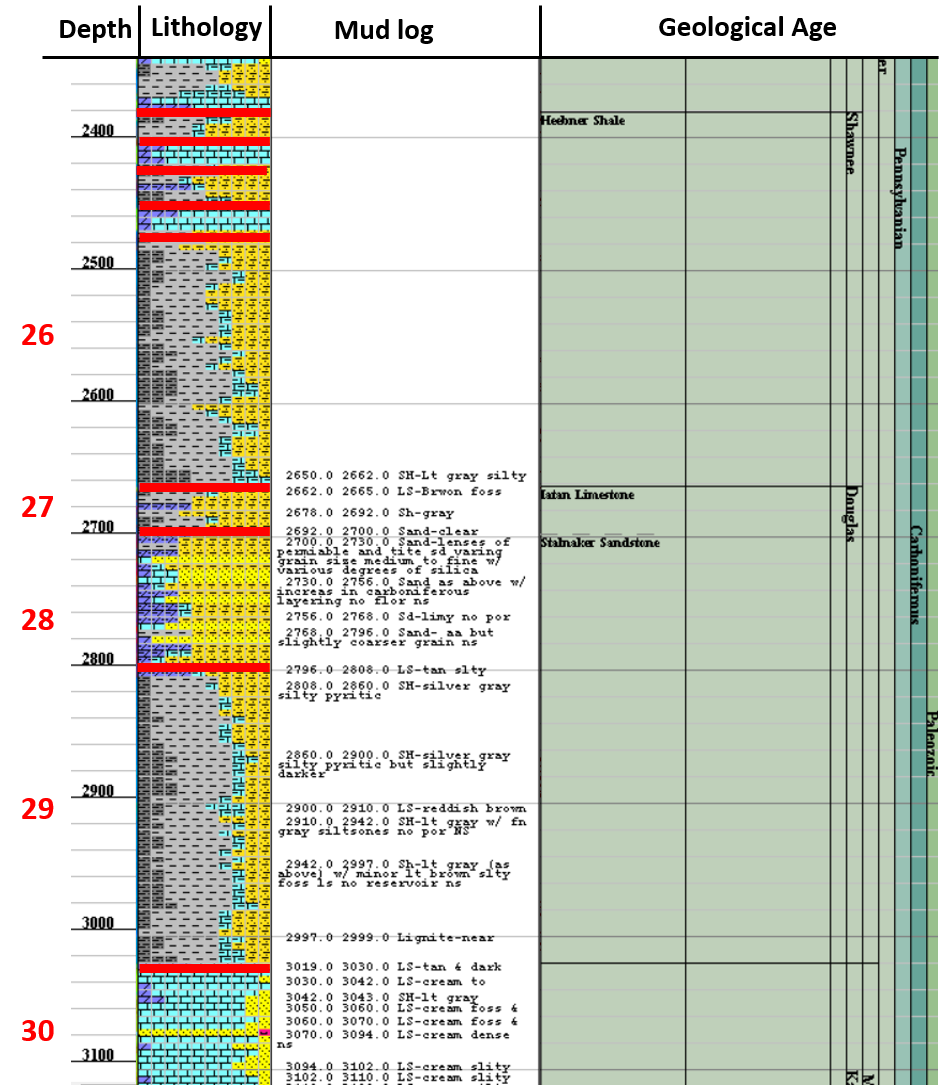}
	\caption[Well 1-28. Well analysis provided by the $CO_{2}$ project]{Well 1-28. Well analysis provided by the $CO_{2}$ project. From the left to the right, we have the depth, the lithology computed, the mud log (brief description of the mineralogy) and the geological age. The  horizontal red lines correspond to the change points found by PELT and the numbers are corresponding to the segmentation of Figure \ref{fig:segmentation_1_28}. The change points are linked to an important change in the lithology.  }
	\label{fig:analysis_1_28}
\end{figure}

In Figure \ref{fig:analysis_1_28}, we display the lithology provided by the Kansas Geological Survey, Oil and Gas Well Database, and also the mud log (description of the mineralogy) with the geological age. We notice that the segmentation algorithm provides layers coherent with the change of lithology given by the experts.

\begin{figure}[!h]
	\centering
		\includegraphics[width=\textwidth]{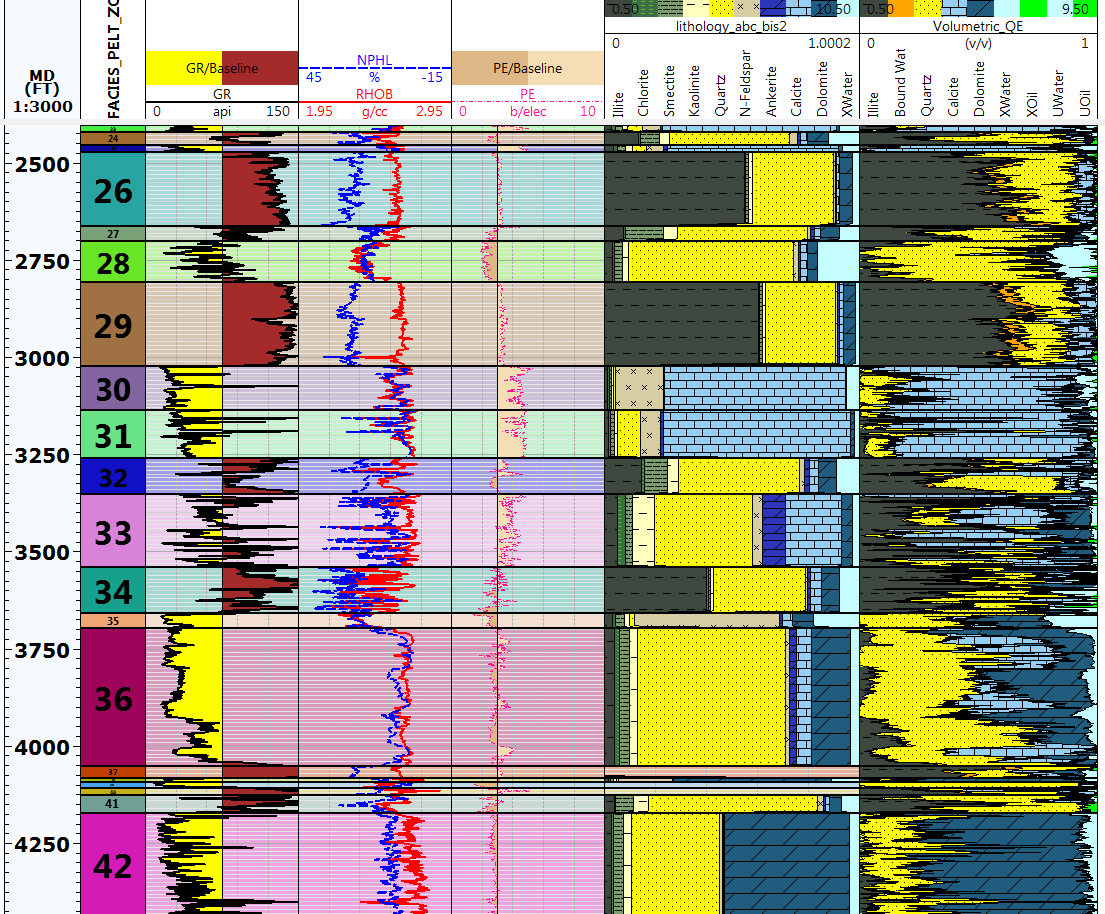}
	\caption[Well 1-28. Results of our methodology]{Well 1-28. Results of our methodology. Track 1: zonation obtained by PELT. Track 2: GR. Track 3: $\rho_{b}$ and $\phi_{N}$. Track 4: $pef$. Track 5: Results of our methodology (displaying the mean lithology of the first hypothesis). Track 6: Lithology obtained by a classical solver using more logs. We notice that the first hypothesis matches quite well the results of the classical inversion even if we are not precise in terms of percentage.}
	\label{fig:results_1_28}
\end{figure}

We apply our methodology on the different segments of PELT using $d=4$ logs (GR, $\rho_{b}$, $\phi_{N}$, $pef$). We use the same lithological model as in the synthetic case with $M=10$ components: Water, 3 carbonates (Ankerite, Dolomite, Calcite), 2 sands (Quartz, N-Feldspar) and 4 shales (Illite, Chlorite, Smectite and Kaolinite). The hyper-parameters of ABC are tuned the following way. We choose $J=10^{6}$ and we set the rejection thresholds for the different logs:
\begin{equation}
  \left\{
      \begin{aligned}
       \delta_{GR} &= 50 \text{ API} \\
       \delta_{\rho_{b}} &= 0.05 \text{ G/$C^{3}$} \\
       \delta_{\phi_{N}} &= 0.03 \text{ V/V} \\
       \delta_{pef} &= 0.2 \text{ b/e} \\  
      \end{aligned}
    \right.
\end{equation}
The  parameter $\delta_{GR}$ is quite important since the values of the Gamma Ray are not necessarily corrected and in a classical mineralogical inversion program, the weight of this log is very light. Because of the rejection method used, it is possible that we do not select in a segment several lithologies if our lithological model does not fit the data. In this case, we set a threshold of an average minimum number of lithology per depth selected by the ABC step to launch the clustering algorithm otherwise we do not give any hypothesis. We fix this quality parameter of the lithological model to 50. In Figure \ref{fig:results_1_28}, we display the main lithological hypothesis found by our method for each zone. In parallel, we performed a classical petrophysical inversion using more logs (8 logs available) and with minerals selected accordingly to the information given by the Kansas study. In general, the first hypothesis matches quite well the main minerals present in the layers. Sometimes, it may be difficult to distinguish minerals from the same family:   Quartz and N-feldspar (layers 30 and 31) or the different clays (layers 27 and 33). Another point is that we are not precise in terms of percentage: in layer 42, we should have around 70\% of dolomite, in our main hypothesis the volume of Dolomite is around 50\%.

\begin{figure}[!h]
	\centering
		\includegraphics[width=\textwidth]{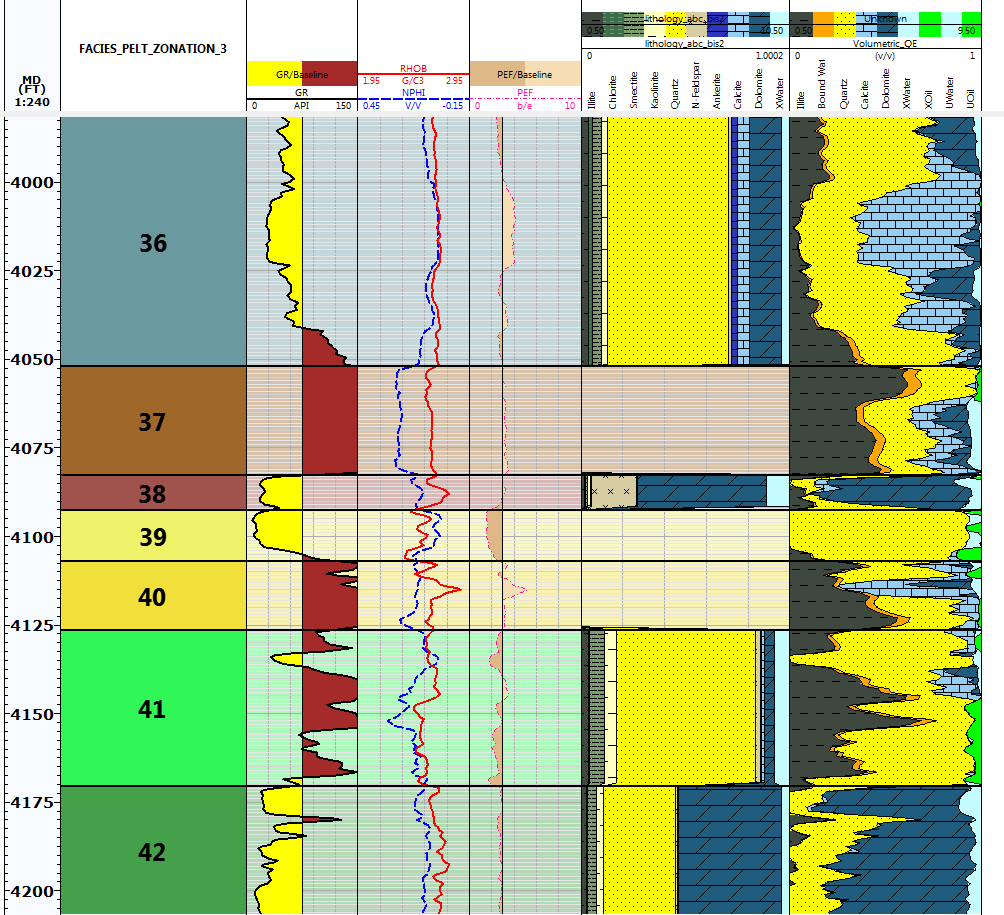}
	\caption[Well 1-28. Results of our methodology focusing on few layers]{Well 1-28. Results of our methodology focusing on few layers. Track 1: zonation obtained by PELT. Track 2: GR. Track 3: $\rho_{b}$ and $\phi_{N}$. Track 4: $pef$. Track 5: Results of our methodology (displaying the mean lithology of the first hypothesis). Track 6: Lithology obtained by a classical solver using more logs. In layers 37, 39 and 40, we do not provide any hypothesis due to the few lithologies selected by ABC step. }
	\label{fig:results_issues_1_28}
\end{figure}

Focusing on the layers  37 to 41 (Figure \ref{fig:results_issues_1_28}), we do not provide any lithological hypothesis for the layers 37, 39 and 40. It is due to the lack of lithology selected by ABC because of a Gamma Ray too high (around 250 API) or a photoelectric factor too low (around 1.6 b/e) which do not fit out lithological model with 10 elements.
Another issue is visible on layer 41: the  main hypothesis is a large volume of sand (around 80\%) and a small volume of clay (around 10\%) whereas the classical inversion program is more balanced (mixed of Quartz and Illite). This kind of error is due to the fact that we select lithologies only in the range of depth where there is a majority of quartz; and we do not pick any lithology in the shaly area because of a high GR. Layer 41 has large variations of its lithology. We must check that the main lithological hypothesis is well distributed on the layer.

\section{Discussion }
We propose a new methodology giving promising results on both synthetic and real examples. The method is fully automated and delivers lithological hypotheses weighted by posterior probabilities. The method needs few logs to produce results, and augmenting the number of logs reduces the uncertainties associated with the hypotheses. We could use other Bayesian inference methods such as Markov chain Monte Carlo (MCMC), but ABC may convince the reader by its simplicity of parametrization and implementation.  However, during the tests, we identified several issues :
\begin{itemize}

\item A problem of identifiability with many minerals in small proportion. We are not able to detect the presence of many minerals of the same family when no log differentiates them. For instance, differentiating carbonate without $pef$ is not easy.

\item Detecting specific minerals such as Kerogen, Pyrite, or Anhydrite may be tricky because adding them into a lithological model always improves the fit and consequently leads to a bias. Our method tends to select them in small proportion. This kind of problem could be solved with better prior on the lithological model.

\item The final selection of minerals is not very clear when we select more minerals than available logs. For instance, if we have a hypothesis with  $M$ components in small proportion (less than 5\% in average) and $M>d+1$, $d$ being the number of logs, we cannot run a classical inversion program to get the lithology depth by depth.

\item We do not integrate hydrocarbons in the lithological sampling model. The problem is that when we introduce oil or gas in a lithological model using the triple combo (GR, $\rho_b$, and $\phi_N$),  they are usually selected even if they are not present. The resistivity equation is needed to determine the difference between water and hydrocarbons, but its parametrization is difficult.

\item  Some clusters selected by HDBSCAN are located on a subsection of the stratum. This problem is due to the segmentation part or the definition of the stratum. Strata with changing or unstable lithologies may yield surprising results: a multiplication of lithological hypotheses covering different subsections of the stratum. 

\item The method may lead to the absence of results because of a lithological sampling model not adapted or noisy data that do not fit in the $\delta_i$ of ABC.

\item The calibration of the HDBSCAN parameter (essentially $m_{pts}$) can be tricky. The default parameter of 5\% of the size of the data seems to work in our tests, but a more complex study is needed.

\end{itemize}

\section{Conclusion}
Some of the issues mentioned above in the Discussion section may be overcome. We give below, as a conclusion, some tracks for potential improvement.

In order to deal with the lack of identifiability and the presence of minerals in small proportion in our  hypotheses, a solution relies in grouping the minerals by family before the clustering phase and apply the clustering on the family. It should reduce the uncertainty on the presence of some mineralogical families.  The size of the stratum could play a role in the determination of the lithology if the logs on the stratum are not homogeneous. Indeed, noisy logs will produce less dense regions in the solution space (lithology) so the clustering algorithm will have difficulties to find lithological hypotheses. Thus, instead of using only a stratum of a single well, applying our method on a global stratum (the same stratum from different wells) should eliminate or differentiate more lithological  hypotheses. For the determination of the water saturation $S_w$ (ratio water/porosity), a possible solution to handle the problem is to first have a rough idea of the porosity using a model without hydrocarbons; and then use the resistivity equation on the first estimation of the porosity to determine the presence of gas or oil. We can iterate the process to improve the precision of the method by adding the selected hydrocarbon in the inference model. It is an imitation of the petrophysicist's workflow. We can also improve the  lithological sampling model with a more complex Bayesian hierarchy by adding prior on the $\alpha$ or the $\delta_i$. For the absence of lithologies selected by ABC, a variant of the method consists in selecting  the $k$ first lithologies generated by the lithological sampling model which minimizes the error between the real logs and the generated one. Choosing $k=500$ gives satisfactory results on synthetic data. Even if the Bayesian inference will always deliver results, the quality of these results can be dubious: the  lithologies can be selected because of only one log (for instance the bulk density). The reliability of the results is an issue.	This concern could eventually be attacked by raising alarms when the data are poor or warning the users when their lithological hypotheses are not appropriate.

\bibliography{abc_article_clean_bib}

\end{document}